\documentclass[%
aip,
rsi,%
amsmath,amssymb,
reprint,%
]{revtex4-1}

\usepackage{dcolumn}
\usepackage{import}
\usepackage{graphicx}
\usepackage{xcolor}
\usepackage{amssymb}
\usepackage{geometry}
\usepackage{csquotes}

\newcommand {\lamu} {\lambda_{\text{u}}}
\newcommand {\E}    {\varepsilon}
\newcommand {\om}   {\omega}

\newcommand {\mum}  {\mu \text{m}}

\newcommand {\Lcr}  {L_{\text{cr}} }

\begin{document}

\title{Channeling and radiation of electrons and positrons in diamond hetero-crystals}

\author{Alexander Pavlov} 
\email{a.pavlov@physics.spbstu.ru}
\affiliation{Peter the Great St.Petersburg Polytechnic University, 
			 Polytechnicheskaya 29, 195251 St.Petersburg Russia}

\author{Andrey Korol}
\affiliation{MBN Research Center, Altenhöferallee 3, 60438 Frankfurt am Main Germany}

\author{Vadim Ivanov}
\affiliation{Peter the Great St.Petersburg Polytechnic University, 
			 Polytechnicheskaya 29, 195251 St.Petersburg Russia}

\author{Andrey Solov'yov}
\altaffiliation[On leave from ]{Ioffe Physical-Technical Institute, St. Petersburg, Russia.}
\affiliation{MBN Research Center, Altenhöferallee 3, 60438 Frankfurt am Main Germany}

\begin{abstract}
We analyze numerically the radiation and channeling properties of 
ultrarelativistic electrons and positrons propagating through a periodically 
bent diamond crystal grown on a straight single-crystal diamond substrate. 
Such systems can be called hetero-crystals and 
they are one of the experimentally realized samples 
for the implementation of crystalline undulators.
We state that in such systems the channeling and radiation properties of 
projectiles are sensitive to the projectile particles energy 
as well as on
the beam propagation direction, 
i.e. on whether the beam of particles enters the crystal from the side of substrate 
or from the side of periodically bent crystal.
The predictions made are important for design 
and practical realization of new crystalline undulators.
\end{abstract}

\maketitle

\section{Introduction}

Development of light sources operating in the photon energy range $E \gg 10$ keV 
is a ambitious goal for modern physics. 
Such light sources can be used in various novel experimental and 
technological applications 
\cite{korol2020crystal}.
%
One of the systems suitable for this task is a crystalline undulator (CU), which 
stands for a periodically bent oriented crystal and a beam of ultra-relativistic particles
that undergo channeling motion \cite{lindhard1965influence}. 
The periodic bending of crystal planes gives rise to a strong, undulator-type CU
radiation (CUR) in the photon energy range 0.1 -- 10 MeV  \cite{korol1998coherent, korol1999photon, book2014channeling}.

Several approaches have been applied to produce periodically bent (PB) crystalline structures.
The most studied system is a strained Si$_{1-x}$Ge$_x$ superlattice in which the concentration $x$ of
the dopant atoms is varied periodically \cite{mikkelsen2000crystalline}.
Such crystals, produced at Aarhus University by means of molecular beam epitaxy, have been 
used in recent channeling experiments with 855 MeV electrons at the MAinzer MIcrotron (MAMI) facility
\cite{backe2013channeling,wistisen2014experimental} 
and with 16 GeV electrons at the SLAC facility \cite{wienands2017channeling}.

\begin{figure*}[ht]
	\centering
	\includegraphics[width=\linewidth]{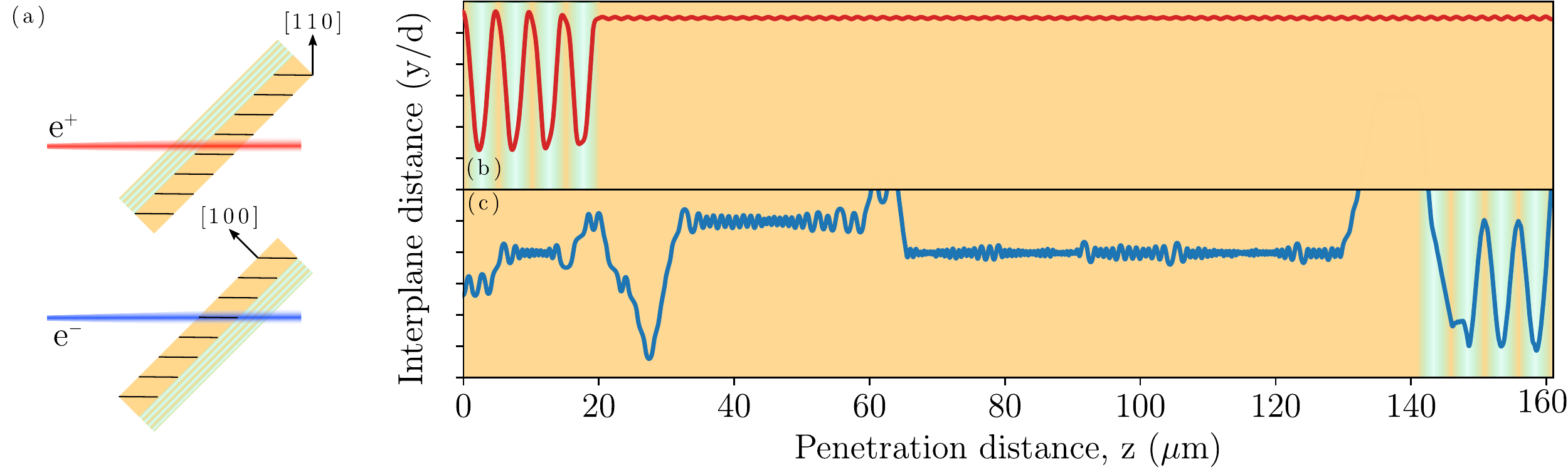}
	\caption{ 
			 (a) Sketch of the crystal geometry.
			 The diamond single crystal is cut with its surface perpendicular 
			 to the $[100]$ direction.
			 The crystal is tilted by $45^\circ$ to orient the  
			 (110) planes along the incident beam, panel (a).
			 The crystal consists of two segments:
			 a straight (S) 141 $\mum$ thick single crystal substrate 
			 and a boron-doped 20 $\mum$ thick periodically bent (PB) segment which accommodates 
			 four bending periods \cite{p38_2016badems}.
			 Gradient shading shows the boron concentration which results in the PB of the (110) planes.
			 Panels (b) and (c) show two possible orientations of the hetero-crystal   
			 with respect to the incident beam.
			 In panel (b) the beam enters the PB segment, in panel (c) -- the S segment. 
			 These two orientations are called "PB-S crystal" and "S-PB crystal", respectively.
			 An exemplary trajectory of a positron channeled through the whole PB-S crystal is shown in
			 panel (b).
			 An \textit{exemplary} trajectory of an electron in the S-PB crystal is presented in panel (c). 
			 Note that several channeling and over-barrier parts the electron's trajectory that are outside the
			 drawing are not shown.
			 }
	\label{fig:geometry}
\end{figure*}

Periodic bending can also be achieved by graded doping during 
synthesis to produce diamond superlattice \cite{tran2017synchrotron}. 
Both boron and nitrogen are soluble in diamond, however, higher concentrations of boron 
can be achieved before extended defects appear 
\cite{tran2017synchrotron,de2007calibration}.
The advantage of a diamond crystal is radiation hardness allowing it
to maintain the lattice integrity in the environment of very intensive 
beams \cite{uggerhoj2005interaction}.
 
Boron-doped diamond layer cannot be separated from a 
straight/unstrained substrate (SC) on which the superlattice is synthesized.
Therefore, unlike Si$_{1-x}$Ge$_x$ superlattice,
a diamond based superlattice has essentially a hetero-crystal structure, 
i.e. it consists of two segments, a straight single diamond crystal substrate and 
a PB layer \cite{p38_2016badems}.

In this paper we present results of the computational analysis 
of channeling and radiation properties in experimentally realized 
\cite{p38_2016badems} diamond based CU, Figure \ref{fig:geometry}.
In our simulations special attention has paid to the analysis of the new effects which appear 
due to the presence of the interface between the straight and PB segments in 
the hetero-crystal.
The experiment has been carried out with the 270-855 MeV electron beams 
\cite{backe2013channeling, p58_2016badems, backe2018channeling}.
For the sake of comparison, the simulations have been carried out for  both electron and positron beams.
The positron beam of the quoted energy range is available at the DA$\rm \Phi$NE acceleration facility 
\cite{backe2011future}.

Geometry of the system is shown in Figure \ref{fig:geometry} (a).
The incident beam can enter the crystal at either PB or straight (S) part.
These two options are shown in panels (b) and (c), respectively.
To distinguish the crystal orientation with respect to the incident beam, 
in the text below we refer to the crystal shown in panel (b) as to the \enquote{PB-S crystal} 
and to the one in panel (c) as to the \enquote{S-PB crystal}.
To illustrate the particle's propagation through the crystal, the selected trajectories of a positron
(red curve, panel (b)) and an electron (blue curve, panel (c)) are shown.

The parameters of the hetero-crystal used in the simulations of 
channeling along the (110) plane correspond to those used in the experiment
 \cite{p38_2016badems}.
 Namely, total thickness in the beam direction is $\Lcr$ = 161~$\mum$
 out of which 141~$\mum$ corresponds to the straight segment and  20~$\mum$ -- to the PB segment.
The cosine bending profile $a \cos(2 \pi z / \lamu)$ was assumed with the 
coordinate $z$ measured along the beam direction.
The bending amplitude and period are $a=2.5$ \AA{} and $\lamu = 5$ $\mu$m, respectively. 
%


In a straight crystal a particle can experience quasi-periodic
channeling oscillations.
In addition to these, a channeling particle in the PB segment is involved in the 
periodic motion due to the periodic bending of a channel.
Spectral distribution of the radiation emitted in a hetero-crystal bears features 
of both types of the oscillatory motion.
For each trajectory simulated spectral distribution of 
electromagnetic radiation has been calculated within the opening angle 
$\theta_0$ = 0.24 $m$rad, which corresponds to one of 
the detector apertures used at MAMI \cite{backe2013channeling}.

Numerical modeling of the channeling and radiation emission processes 
was performed using \textsc{MBN~Explorer} computational software
\cite{solov2012mesobionano}.
By means of its channeling module \cite{sushko2013simulation} it is 
possible to simulate motion of ultra-relativistic particles 
in different environments, including the crystalline ones.
%
%
The method of all-atom relativistic molecular dynamics is described
in great details in Refs. \cite{sushko2013simulation, shen2018channeling}.
%

\section{Results and discussion for positrons}

\begin{figure}[htb]
	\centering
	\includegraphics[width=\linewidth]{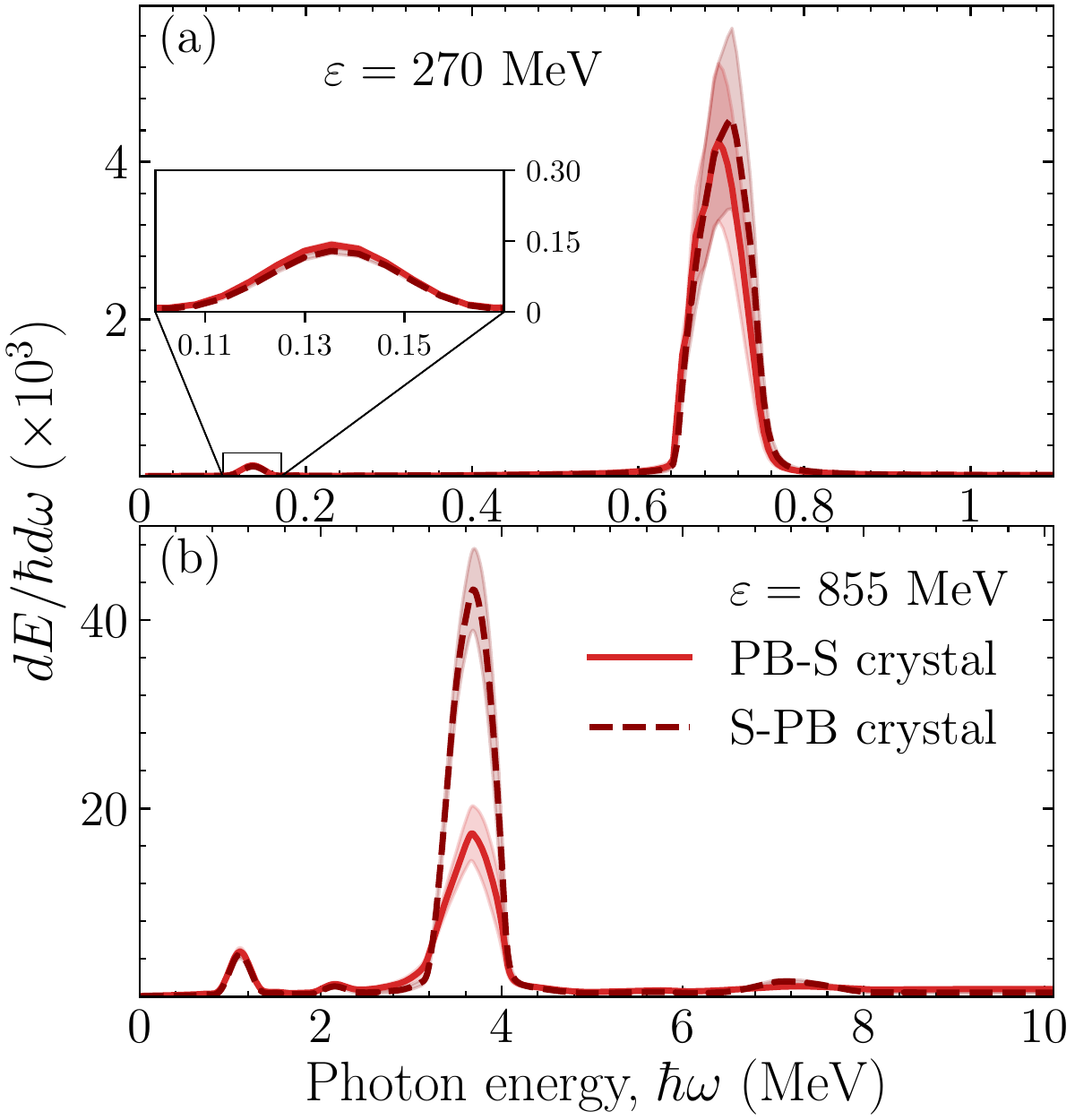}
	\caption{Radiation emission spectra produced by $\E = 270$
			and 855 MeV positrons in $\Lcr =$ 161 $\mum$ 
			diamond(110) hetero-crystals.
			The collection angle was set to $\theta_0 = 0.24$ $m$rad.
			(a) Spectra by 270 MeV positrons propagated 
			in PB-S and S-PB crystals. 
			(the inset shows the $\times$6 magnified CUR peak) 
			(b) same as (a) but for 855 MeV positrons.
			Shading indicates the statistical error due to the 
			finite number of simulated trajectories.
			For the sake of comparison we quote the intensities 
			of the background incoherent bremsstrahlung estimated 
			within the Bethe-Heitler approximation: 
			$2.9 \times 10^{-6}$ and $2.5 \times 10^{-5}$
			for $\E = $270 and 855 MeV, respectively.} 
	\label{fig:p_270_855_spectra}
\end{figure}

Let us now analyze the case of positron channeling in hetero-crystals.
In the planar channeling regime, a charged projectile moves along 
a crystallographic planes experiencing a collective electrostatic 
field of the lattice atoms \cite{lindhard1965influence}.
For positrons, the atomic field is repulsive, so that the particle 
channels in between two adjacent crystalline planes. 
In this case, nearly harmonic channeling oscillations give rise to 
narrow photon emission lines.

Figure \ref{fig:p_270_855_spectra} presents the spectra calculated 
for the 270 and 855 MeV projectiles.
The spectra consist of two main parts: the CUR (peak at lower energies)
and ChR (peak at higher energies).
The CUR radiation emits from the PB segment while 
the ChR can be generated in the PB and straight segments of the crystal.

Figure \ref{fig:p_270_855_spectra} (a) shows the results for 270 MeV positrons.
The spectra are dominated by peak of ChR 
(strong peak at $\hbar \om \approx 0.7$ MeV), CUR reveals itself as 
a small bump (note the insert in Figure \ref{fig:p_270_855_spectra} (a)) 
in low energetic part of the spectra ($\hbar \om \approx 0.13$ MeV). 
It is easy to notice that for $\E =$ 270 MeV electrons 
the spectral densities of CUR and ChR for PB-S and S-PB crystals 
are deviate within margin of statistical error.

\begin{figure}[htb]
	\centering
	\includegraphics[width=\linewidth]{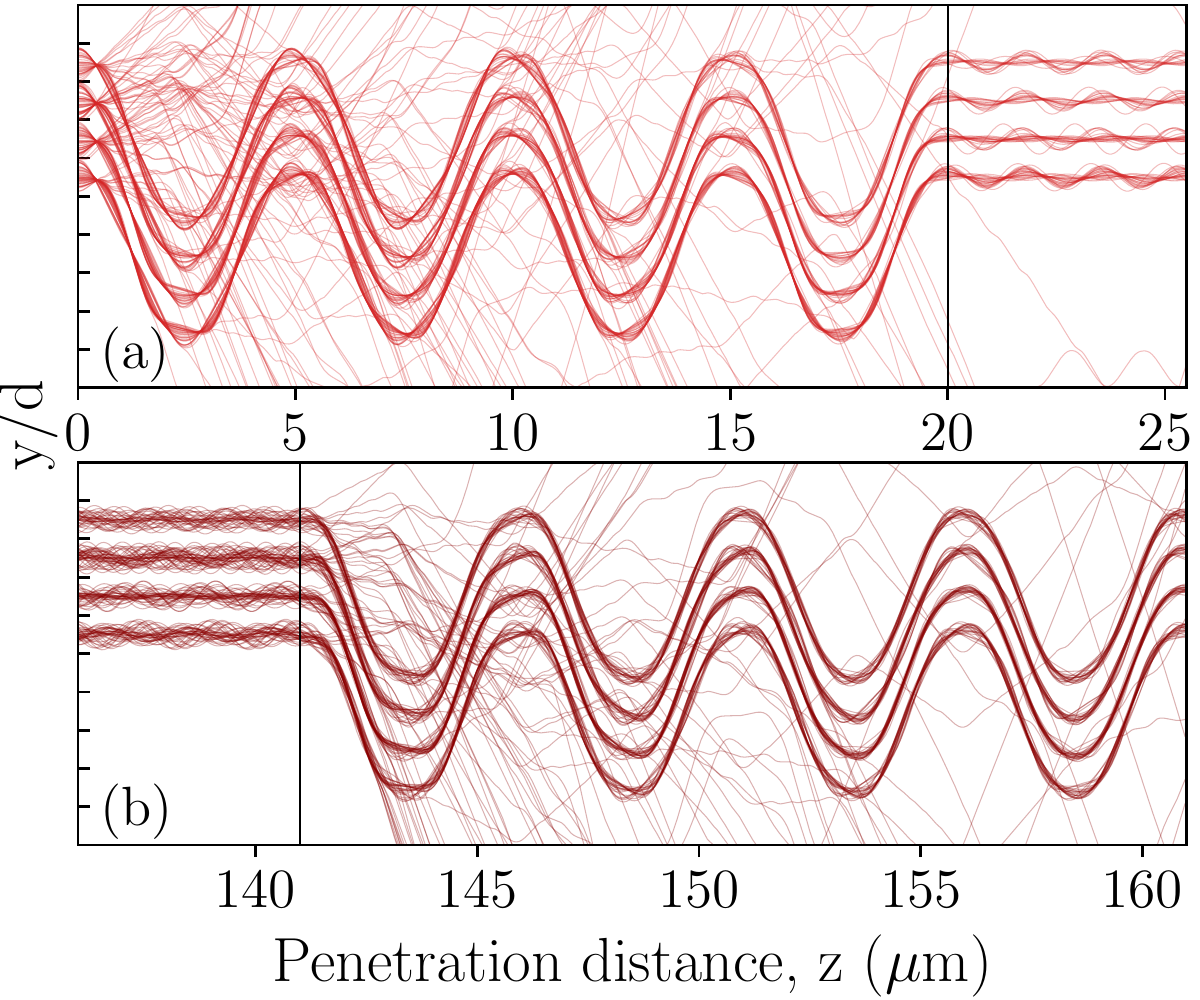}
	\caption{2D projections of several \textit{exemplary} 
			 trajectories of $\E = 855$ MeV positron 
			 propagated in $\Lcr =$ 161 $\mum$ diamond hetero-crystal.  
			 (a) trajectories of positrons which propagate in PB-S crystal 
			 To highlight the processes which occur on the interface between 
			 PBC and SC the trajectories was zoomed 
			 into the window with z = 0 -- 25 $\mum$. 
			 (b) same as (a), but in S-PB crystal, 
			 the zoom window is set to 136 -- 161 $\mum$.
			 The vertical black lines in (a) and (b) indicate the position of interface 
			 between PB and straight segments of the crystal, 
			 20 $\mum$ and 141 $\mum$ correspondingly.} 
	\label{fig:p_855_trj}
\end{figure}

For 855 MeV positrons spectra are shown in Figure \ref{fig:p_270_855_spectra} (b).
The spectra consist of two main peaks: 
the peak of CUR around $\hbar \om \approx 1.1$ MeV and 
the peak of ChR around $\hbar \om \approx 3.6$ MeV.
One can notice small bumps around $\hbar \om \approx 2.2$ MeV and 
$\hbar \om \approx 7.2$ MeV which is second harmonics of CUR and ChR respectively.
As well as for $\E = $ 270 MeV positrons intensities of CUR 
are the for two types of crystal, 
but spectral densities of ChR in S-PB is $\approx 2$ times large than this in PB-S. 
In case of 855 MeV positrons the difference in the intensities between 
CUR and ChR should be even less pronounced in the experiments,
where usually, not the spectral density $dE/\hbar dw$ 
is measured, but the number of photons 
$(1 / \hbar d \om) dE/\hbar d \om$ with certain energy. 

\begin{figure*}[t]
	\centering
	\includegraphics[width=\linewidth]{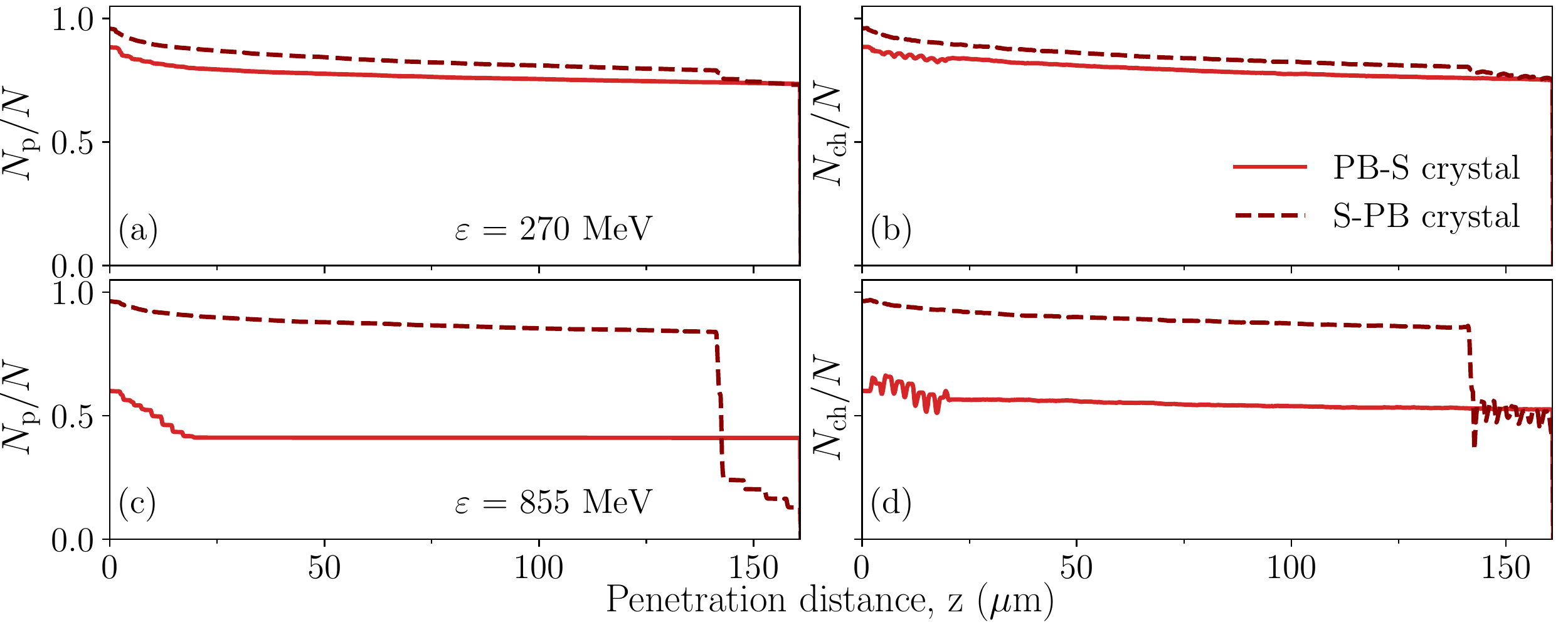}
	\caption{Fraction of channeling positrons in 
			 $\Lcr =$ 161 $\mum$ diamond(110) hetero-crystals.
			 (a) primary fractions of the particles for $\E$ = 270 MeV positrons, 
			 (b) fractions with account of the re-channeling for positrons with energy 
			 $\E$ = 270 MeV,
			 (c) - (d) same as (a) - (b) but for positrons with energy $\E$ = 855 MeV} 
	\label{fig:p_270_855_length}
\end{figure*}

In order to analyze difference in the radiation spectra,
let us plot the trajectories of positrons. 
Figure \ref{fig:p_855_trj} presents the \textit{exemplary} 
trajectories of 855 MeV positrons.
Figure \ref{fig:p_855_trj}(a) shows parts of the trajectories 
of positrons which channels in PB-S crystals.
In that case positrons firstly enter through the PB crystal and than 
penetrate through the interface to the SC.
For $\E =$ 855 MeV positrons the amplitude of channeling oscillations 
is strongly suppressed, since the potential barrier is reduced due to centrifugal
force \cite{pavlov2019interplay,pavlov2020channeling}.
This also results in strong suppression of ChR for high energetic particles
in PB diamond crystals.
Because of that, the positrons which are propagating in the PBC experience strong 
dechanneling in the parts of the crystal with large curvature of its planes 
where the centrifugal force acting on the channeled positron is maximal.
However, the opposite process, re-channleing, can occur in the segments of the crystal
with small curvature.
Re-channeling results in effective increase in total channeling length of the particles.
Positrons, which are captured to the channel in PB segment then penetrate 
to the straight segment without dechanneling. 
Propagating through the interface they retain the amplitude 
of their channeling oscillations.

In opposite situation when the positrons propagate in S-PB crystal 
(Figure \ref{fig:p_855_trj} (b)) they first come through the straight segment 
and then penetrate to the PB segment.
However, positrons can channel in the straight segment with transverse
energies higher than in the PB segment.
This results in higher amplitude of channeling oscillations in straight segment.
As a result strong dechanneling occurs on the interface between the straight
and PB segments of the crystal (Note: the dechanneling in the region between 
140 and 145 $\mum$ in Figure \ref{fig:p_855_trj} (b)).

To illustrate the channeling properties of positrons, 
we plot (Figure \ref{fig:p_270_855_length}) the dependence of primary 
fraction and fraction of channeled particles
with account of re-channeling as a functions of penetration distance $z$.
These dependencies can be used to analyse the intensities of CUR and ChR
in PB-S and S-PB crystals.
Since, the intensity of radiation due to periodic motion 
is proportional to the number of particles participating in quasi periodic motion
and square of the amplitude of 
corresponding oscillation \cite{pavlov2019interplay,pavlov2020channeling}.

For $\E =$ 270 MeV positrons the dependencies of primary fraction, 
Figure \ref{fig:p_270_855_length} (a),
and fraction with account of re-channeling Figure \ref{fig:p_270_855_length} (b)
are almost identical for both types of crystal.
The acceptance in case of S-PB crystal is higher than for PB-S crystal,
due to the centrifugal.
Since, the amplitude of the periodic bending $a$ is equal for both 
types of crystal and the number of particle involved in the channeling motion
in PB parts of the crystals are approximately same the intensity of CUR should be
the same (see Figure \ref{fig:p_270_855_spectra}).
Same is true for the intensity of ChR.
The centrifugal force is small and as result, change in channeling amplitudes 
is also small \cite{pavlov2019interplay}.
Thus the number of channeled particles in PB-S and S-PB crystals are comparable.

For $\E =$ 855 MeV the situation is different, 
due to the increase of the centrifugal force in the PB parts 
of the crystals  the number of particles involved into the channeling motion
drops as well as alternates the amplitude of channeling oscillations.
The primary fraction of positrons in PB-S crystal 
decrease steadily drops every time particles 
propagate through the segments of the crystal with high curvature
(note the step-like dependence in first 20 $\mum$ in 
Figure \ref{fig:p_270_855_length} (c)).
However, in the straight segment of the crystal this dependence remains constant,
since the centrifugal force caused by the crystal bending is absent in the straight
segment of the crystal.
Account for re-channeling gives rise to oscillations of the number 
of channeled particles in PB segment and 
the rise of number of channeling particles in straight segment.

For $\E =$ 855 MeV positrons propagating behavior of the dependencies is different.
In the case of S-PB crystal the value of acceptance is higher and no significant re-channeling
occurs in the straight part of the crystal.
However, the drastic drop in number of primary particles appears at the interface.
This happens due to appearance of the centrifugal force in the PB part of the crystal.
The number of channeling particles with account of re-channeling 
become approximately equal to the  number of particles 
in the channeling regime in case of PB-S crystal 
(see Figure \ref{fig:p_270_855_length} (d)).

As a result of such behavior of channeled positrons,
the intensity of CUR are approximately same.
But, the intensity of ChR for the PB-S crystal should be about 2 times 
smaller than for the S-PB crystal (see Figure \ref{fig:p_270_855_spectra} (b)).

To conclude, two main effects are observed for positrons in 
two types of crystals: 
1. For the two cases considered ChR intensities are same 
within margin of errors for $\E =$ 270 MeV 
and differs at least two times for $\E = 855$ MeV
positrons.
2. CUR intensities are the same for both types of crystals.
They can be explained by the presence of the centrifugal force
in the PB segments of the crystal.

\section{Results and discussion for electrons}

Let us now return to the analysis of electron channeling in hetero-crystals.
Electrons, in contrast to positrons, move around crystalline chains.
This results in increase of the number of hard collisions with bulk constituents
and thus the dechanneling/re-channeling rates. 

\begin{figure}[ht]
	\centering
	\includegraphics[width=\linewidth]{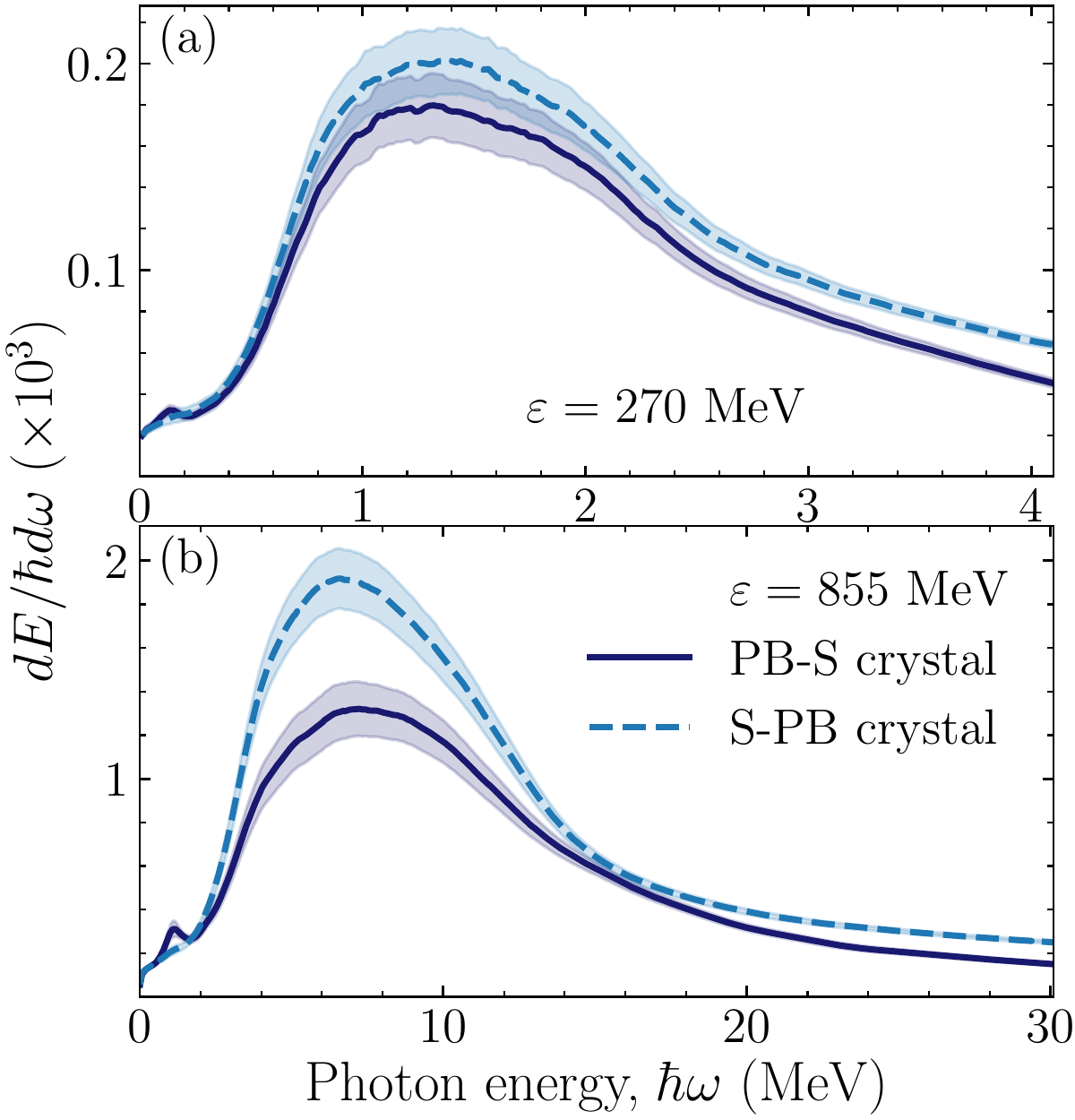}
	\caption{Same as Figure \ref{fig:p_270_855_spectra} but in case of electrons.} 
	\label{fig:e_270_855_spectra}
\end{figure}

The anharmonicity of the interaction potential between ultra-relativistic
electrons and lattice atoms results in significant  
broadening of the peaks.
The examples of the spectra are shown in Figure \ref{fig:e_270_855_spectra}.

In Figure \ref{fig:e_270_855_spectra}(a) the results for $\E = 270$ MeV 
electrons are presented.
For the given electron energy the intensities of ChR for the two geometries coincide 
within the margin of statistical error.
For PB-S crystal the small bump corresponding to the CUR arises in the radiation spectra
around $\hbar \om \approx 0.13$ MeV.
In the case of S-PB crystal this peak is nearly absent.

For $\E = 855$ MeV electrons the intensities of ChR differ approximately $\approx 1.5$ times.  
The CUR radiation for 855 MeV electrons in the PB-S crystal reveals itself 
as a peak at the photon energy $\hbar \om \approx 0.7$
This peak is much more pronounced $\E = 855$ MeV electrons than for 270 MeV 
and nearly absent for S-PB crystal.

\begin{figure*}[ht]
	\centering
	\includegraphics[width=\linewidth]{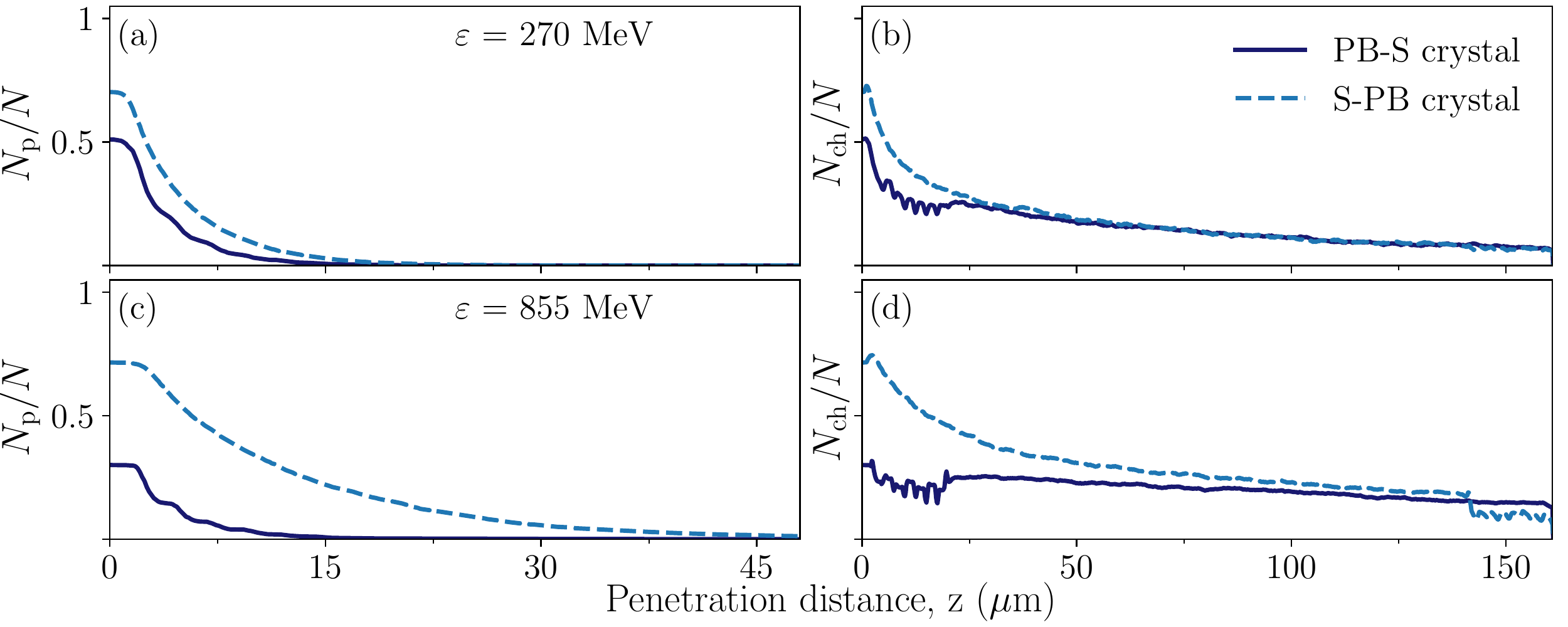}
	\caption{Same as Figure \ref{fig:p_270_855_length} but in case of electrons.} 
	\label{fig:e_270_855_length}
\end{figure*}

In order to understand these behaviors, one can plot the dependencies of 
fraction of accepted electrons (Figures \ref{fig:e_270_855_length}) (a), (c))
and fraction of electrons with account of re-channeling
(Figures \ref{fig:e_270_855_length} (b), (d)) upon the penetration distance.
As for positrons, the value of $N_p/N$ at ($z = 0$) corresponds to the acceptance value.
For PB-S the acceptance is determined by the PB part of the crystal. 
Thus it is reduced by by the presence of the centrifugal force.
For the S-PB crystal the acceptance is the same as in the straight crystal.

Compared to positrons, electrons have significantly shorter dechanneling lengths,
this leads to the fact that the number of primary fraction of channeled 
electrons practically dies out for crystal thicknesses greater than 50 $\mum$, 
both for PB-S and for S-PB crystals.
At both 270 and 855 MeV CUR emission requires a particle to pass at least one full
channel period CU.
If channeling length becomes shorter than than half of CU period 
photon emission by electrons becomes similar to synchrotron radiation.
Taking into account the dynamics of dechanneling/re-channeling of electrons in 
PBC \cite{korol2017channeling}, the main contribution to CUR is produced 
by particles accepted into the channeling regime \cite{pavlov2019interplay}.
This explains why CUR radiation arises in spectra mostly in the case of PB-S crystal.
Also, it explains why CUR peak is more pronounced for 855 MeV electrons
than for 270 MeV ones.

ChR intensity is proportional to the number of particles involved
in channeling the motion.
The difference in number of channeled electrons with energy 270 MeV
in the two cases (Figure \ref{fig:e_270_855_length} (b))
manifests itself only at the first 20 $\mum$ of the crystals.
Beyond this region the difference is negligible.
This results in a small difference in the ChR intensities for
270 MeV of electrons in two cases consider which falls within the margin of error.
For $ \varepsilon = 855 $ MeV the difference in the number of channeled electrons
significant, leading to a greater difference in spectral
ChR intensities in Figure \ref{fig:e_270_855_spectra} (b).

To conclude this section let us state:
1. For the two cases considered ChR intensities are the same 
within margin of errors for $\E =$ 270 MeV 
and become different for the $\E =$ 855 MeV electrons. 
2. CUR vanishes in the case of S-PB crystals and is present for PB-S crystals. 
Such behavior, as for positrons, can be explained by the
centrifugal force acting on the electrons in the PB segments of the crystals. 
The manifestation of CUR is also determined by relatively short 
dechanneling length of electrons with respect to positrons.

\section{Conclusions}

\begin{figure}[ht]
	\centering
	\includegraphics[width=\linewidth]{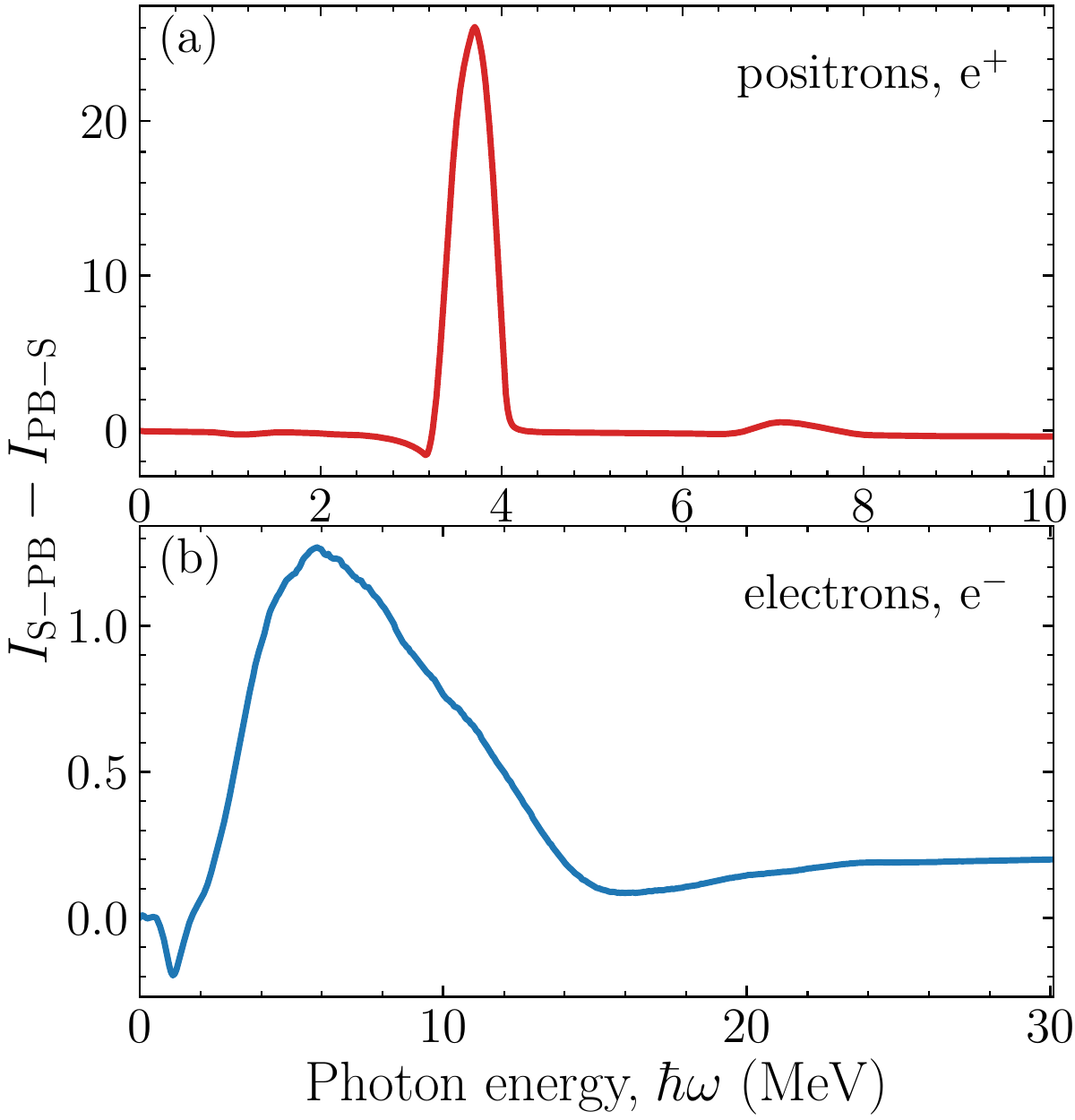}
	\caption{Difference between radiation intensities for two propagation direction 
			for 855 MeV (a) positrons and (b) electrons. 
			The data is taken from Figure \ref{fig:p_270_855_spectra} (b) and 
			Figure \ref{fig:e_270_855_spectra} (b) respectively.
			$I_{\text{S-PB}} $ stands 
			for the spectral densities of radiation $dE / \hbar d \omega$
			for particles propagating in S-PB crystal, 
			$I_{\text{PB-S}}$ same but for PB-S crystal.}   
	\label{fig:ep_855_difference}
\end{figure}

In summary, the channeling and radiation phenomena for 270 and 855 MeV 
positrons and electrons in oriented diamond hetero-crystals were simulated 
by means of all-atom relativistic molecular dynamics. 
We predict the radiation spectra for the two orientations (PB-S and S-PB) 
of crystals with respect to the beam.
In the case of  positrons the peaks of CUR are clearly distinguishable from the background 
and have comparable intensity with respect to the ChR peaks.
However, in the case of electrons, 
the CUR peaks observation over the broad peak of ChR becomes a challenging task, 
especially at low energies.
This prediction opens a possibility for the experimental detection 
of CUR in the PB structures grown on a substrate.

One possible way to analyse experimental data for such systems
is to measure the radiation spectra for particles propagating in PB-S and S-PB
crystals and analyse their ratio.
An example of such analysis is shown in Figure \ref{fig:ep_855_difference}.
In this case not only the enhancement of radiation in the CUR desired region,
but als the difference in channeling radiation intensities can be a fingerprint 
of the PB segment inside the crystal. 

Finally, the analysis performed demonstrates that usage of 
PBC with a straight substrate does not provide advantages for the CUR production.
However, in the cases when a high-quality PB crystals 
can only be produced as segments of hetero-crystals
practical realization of CUs based on such crystal with high quality 
positron beams is a feasible task.
Electron beams can be used for probing the quality 
of PB segments of hetero-crystals.

\begin{acknowledgments}
	The work was supported in part 
	by the DFG (Project No. 413220201)
	and by the N-LIGHT Project 
	within the H2020-MSCA-RISE-2019 call (GA 872196). 
	We acknowledge the Supercomputing Center of 
	Peter the Great Saint-Petersburg Polytechnic University 
	(SPbPU) for providing the opportunities to 
	carry out large-scale simulations.
	We are grateful to Hartmut Backe and Werner Lauth (University of Mainz) for
	useful discussions, 
	to Rostislav Ryabov (SPbPU) for careful reading of the manuscript.	
\end{acknowledgments}

\bibliography{lib_chan}

\end{document}